\documentstyle[12pt]{article}

\input{psfig.tex}

\renewcommand{\baselinestretch}{1.5}

\newcommand{\NPB}{Nucl. Phys. {\bf B}}
\newcommand{\PRD}{Phys. Rev. {\bf D}}
\newcommand{\PLB}{Phys. Lett. {\bf B}}

\begin{document}

\begin{flushright}
June 1998, BROWN-HET-1130, hep-th/9806125 \\ 
\end{flushright}     
\begin{center}
{\Large 
{\bf Evaluation of Glueball Masses From Supergravity
\footnote{
supported by 
DOE under grant DE-FG0291ER40688-Task A; RdMK is supported by a South Africa
 FRD Postdoctoral Bursary;  antal,nunes,rdm@het.brown.edu, 
mm@barus.physics.brown.edu}}}\\
Robert de Mello Koc${\rm h}^{\dagger}$, Antal Jevick${\rm i}^{\dagger}$, 
Mihail Mihailesc${\rm u}^{\dagger}$ and Jo\~ao P. Nune${\rm s}^{\dagger ,*}$\\
Department of Physic${\rm s}^{\dagger}$\\
Brown University\\
Providence RI 02912, USA \\
and\\
Departamento de Matem\'atic${\rm a}^{*}$ \\
Instituto Superior T\'ecnico \\
Av.Rovisco Pais, 1096 Lisboa Codex, Portugal \\
{\bf Abstract} 
\end{center}
In the framework of the conjectured duality 
relation between large $N$ gauge theory and supergravity the spectra 
of masses in large $N$ gauge theory can be determined by solving certain
eigenvalue problems in supergravity. In this paper 
we study the eigenmass problem given by Witten as a possible approximation 
for masses in QCD without supersymmetry. We place a particular
emphasis on the treatment of the horizon and related boundary conditions.
We construct exact expressions for the analytic expansions 
of the wave functions both at the horizon and at infinity and show that 
requiring smoothness at the 
horizon and normalizability gives a well defined eigenvalue problem. We show 
for example that there are no smooth solutions with vanishing derivative at 
the horizon. 
The mass eigenvalues up to $m^{2}=1000$ corresponding to smooth normalizable 
wave functions are presented. We comment on the relation of our work with 
the results found in a recent paper by Cs\'aki et al., hep-th/9806021, which 
addresses the same problem.

\newpage
\section{Introduction}

The problem of solving QCD in the nonperturbative large $N$ limit has been
outstanding for several decades \cite{'tHooft,Pol}. It has been suspected
that the answer to this question will come from string theory. Recently a 
very interesting proposal \cite{JM} has been introduced and further 
explored \cite{JM2,EW2,EW1} which involves a relationship between large $N$ 
super Yang-Mills theory and $AdS$ supergravity \cite{GKP}-\cite{GKT}. 
This correspondence which was first investigated in studies of 3-branes  
gives the possibility of studying large $N$ properties of Yang-Mills 
theories using classical supergravity \cite{Kle}. The later is 
expected to give results 
that should be valid for the strongly coupled gauge theory. At present time 
comparison of the two theories was done for operators and correlators 
protected by supersymmetry \cite{FMMR}-\cite{LMRS}.
Other quantities like the entropy or Wilson loops represent predictions of the
conjecture \cite{JM}-\cite{GKT}. For general systems involving p-branes a 
notion of generalized conformal symmetry was found in \cite{JY}.
One can expect that a similar correspondence holds also in  theories without 
supersymmetry and ultimately in QCD. Witten has presented such an extension
where proprties of finite temperature Yang-Mills theories are to be
computed using $AdS$ black hole backgrounds in gravity \cite{EW1}.

According to Witten's generalization of the conjecture by Maldacena in 
\cite{JM}, in order to study ${\cal N}=4$ super Yang-Mills theory at large 
$N$, high temperature and strong t'Hooft coupling, 
one should consider the Euclidean Schwarzchild black hole solution in 
$AdS_{5}\times S^{5}$ space-time in the limit where the black hole 
mass is large \cite{EW1}. In this limit the metric can be written as
\begin{equation}
ds^{2}= (\frac{r^{2}}{b^{2}}-\frac{b^{2}}{r^{2}})d\tau^{2}+
\frac{dr^{2}}{(\frac{r^{2}}{b^{2}}-\frac{b^{2}}{r^{2}})}
+r^{2}\sum_{i=1}^{3}dx_{i}^{2}+b^{2}d\Omega_{5}^{2}
\label{1.1}
\end{equation}
where $d\Omega_{5}^{2}$ is the round metric on $S^{5}$, $r=b$ is the horizon 
radius and the coordinate $\tau$ is the Euclideanized periodic time coordinate.
This metric is obtained as a solution to the type IIB supergravity equations
of motion following from the $\gamma \rightarrow 0$ limit of the action 
\begin{equation}
S=-\frac{1}{16\pi G_{10}}\int d^{10}x\sqrt{g} (R-\frac{1}{2}(\partial\phi)^{2}
+\cdots + \gamma \exp (-\frac{3}{2}\phi)W + \cdots)
\label{1.2}
\end{equation}
where the order $\gamma =1/8 \zeta (3) \alpha^{'3}$ terms contain the first 
string corrections to supergravity and where $W$ is a certain combination of 
terms quartic in the Weyl tensor \cite{BG,GKT}. The inclusion of stringy 
$\alpha^{'}$ corrections corresponds to including strong coupling expansion 
corrections in the gauge theory \cite{JM,EW2,EW1}.

In this paper we would like to study the proposal in \cite{EW1} for the 
supergravity calculation of the mass gap in QCD.
In the next section we will examine the equations of motion for free scalar 
field propagation on $AdS$ black hole backgrounds by rewriting them in
the form of an Hamiltonian problem. We will then address the problem of
the behaviour of the wave function at the horizon. In section 3 we present
our exact results for the wave functions and show that there are no 
normalizable smooth solutions with vanishing derivative at the horizon. Using 
the exact form of the solutions we then exhibit the glueball mass eigenvalues 
predicted by supergravity. Finally, in section 4 we close with some 
conclusions.

\section{Black Hole Backgrounds}

According to Witten  the equations for free field propagation, 
$\partial_{\mu}(\sqrt{g}g^{\mu\nu}\partial_{\nu}\eta)=0$ for a scalar field,
in the five-dimensional space-time described by the 
first terms of (\ref{1.1}) with $\tau$ compactified on $S^{1}$ should 
give glueball masses for $QCD_{3}$. Similarly a computation in an 
$AdS_{7}$ black hole background 
is expected to be of relevance for glueball masses in $QCD_{4}$ \cite{EW1}. 
One should look for solutions behaving like (static) 
plane waves along the $x_{i}$ directions $\phi\sim\eta (r)e^{ik\cdot x}$ and 
then demanding normalizability and regularity of the behaviour of $\eta (r)$ 
at $r=b$ and $r=\infty $ will select only certain allowed discrete values for 
$m^{2}=-k^{2}$. These values of $m^{2}$  are then interpreted as particle 
masses in the three-dimensional world parametrized by the $x_{i}$. 
To study the corrections to these masses in the strong coupling expansion one
then should work with the $O(\gamma)$ corrections to the background metric
(\ref{1.1}) and to the dilaton field. To $O(\gamma)$ and for the purpose of 
computing mass corrections, it is consistent to take a classical solution with 
a vanishing dilaton field. The $O(\gamma )$ correction to the metric was 
found in \cite{GKT} and one uses it to compute the $O(\gamma )$ corrections 
to the glueball masses.

Consider the general metric for the $AdS_{n+1}$ black hole in the large mass 
limit \cite{EW1}
\begin{equation}
ds^{2}=(\frac{r^{2}}{b^{2}}-\frac{b^{n-2}}{r^{n-2}})d\tau^{2}+
\frac{dr^{2}}{(\frac{r^{2}}{b^{2}}-\frac{b^{n-2}}{r^{n-2}})}
+r^{2}\sum_{i=1}^{n-1}dx_{i}^{2}.
\label{new.1}
\end{equation}

The equation of motion for a free scalar field of the form $\phi\sim\eta (r)
e^{ik\cdot x}$ is then given by
\begin{equation}
\partial_{r}(r^{n-1}(\frac{r^{2}}{b^{2}}-\frac{b^{n-2}}{r^{n-2}})\partial_{r}
\eta )+r^{n-3}m^{2}\eta =0
\label{new.2}
\end{equation}
where $m^{2}=-k^{2}$ is the $(n-1)$-dimensional mass. Consider the measure 
coming from the metric (\ref{new.1}) above (we set $b=1$ for the remainder 
of this section)
\begin{equation}
<\eta |\eta>=\int_{1}^{\infty}dr r^{n-1}\eta(r)\eta^{*}(r).
\label{new.3}
\end{equation}
In order to trivialize the measure we can take a new variable 
$y=r^{\frac{n}{2}}$ for which with $\Phi (y)=y^{\frac{1}{2}}\eta (y)$ 
the equation becomes
\begin{equation}
\partial_{y}((y^{2}-1)\partial_{y}\Phi )+
(-\frac{1}{4}(3+y^{-2})+\frac{4m^{2}}{n^{2}}y^{-\frac{4}{n}})\Phi =0
\label{new.4}
\end{equation}
Integrating (\ref{new.4}) from the horizon to $\infty$ against $\Phi^{*}(y)$ 
and integrating by parts assuming normalizability and smoothness we obtain a 
bound on the possible values of $m^{2}$. For example, for $n=4$ we find that 
$m^{2}>4$. To eliminate the first derivative term in (\ref{new.2}) and to write
that equation in terms of an Hamiltonian problem we now take 
$y=r^{\frac{n}{2}}=\cosh (w)$ and redefine $A(w)=\sinh (2w)^{\frac{1}{2}}
\eta(w)$. This gives 
\begin{equation}
\frac{1}{2}\partial_{w}^{2}A(w) -V(w)A(w)=0
\label{new.5}
\end{equation}
where the potential is now given by
\begin{equation}
2V(w)=1-\sinh (2w)^{-2} - \frac{4}{n^{2}}m^{2}\cosh (w)^{-\frac{4}{n}}.
\label{new.6}
\end{equation}
We are interested in the wave function $A(w)$ 
for the zero eigenvalue of (\ref{new.5}). If we expand the potential around
the horizon $w=0$ we obtain $V(w)=-1/(8w^{2})+(2/3-(2m^{2})/n^{2})+
((4m^{2})/n^{3}-2/15)w^{2}+O(w^{4})$. The harmonic oscillator perturbed by
a potential of the form $\lambda(1/w^{2})$ was examined in \cite{L}. Our 
potential corresponds precisely to the limiting case $\lambda =-1/8$ in that 
reference beyond which the Hamiltonian is not bounded below.
The indicial equation for (\ref{new.5}) with the potential expanded about
the horizon $w=0$, will have a double root $\frac{1}{2}$. Therefore, near 
the horizon we will have the behaviours $A(w)\sim w^{\frac{1}{2}}$ and $A(w)
\sim w^{\frac{1}{2}}\log (w)$ for the two independent solutions of 
(\ref{new.5}). Both solutions are normalizable near $w=0$  and we also have 
normalizable density of probability currents at the horizon of the form 
$J(w)\sim A(w)\partial_{w}A(w)
\sim {\rm const.}\;\;{\rm or}\;\;\sim\log^{2}(w)$.
The two solutions give wave functions $\eta(r)$ for (\ref{new.2}) which 
behave near the horizon like $\eta (w)\sim {\rm const.}\;\;{\rm or}\;\;
\eta (w)\sim\log (w)$. The first derivatives then become $d\eta /dr\sim
{\rm const.}\;\;{\rm or}\;\;d\eta /dr\sim {\rm const.}/w$. We therefore 
expect that the Neumann boundary condition may never be attained at the 
horizon for
a regular solution. Indeed we note that we have a potential which is singular 
at the horizon and that it could be expected that it is not possible to 
demand Dirichlet or Neumann boundary conditions there and as we will see 
this is what happens 
in our case. It would be interesting from this general point of view to 
understand if possible tunneling effects could contribute in a small amount 
to the values of the eigenmasses.
Our solutions of equation (\ref{new.2}) which we will 
present in the next section are consistent with the above behaviour.

To formulate the eigenvalue problem, 
one fixes the behaviour at $\infty$ such that the solution is normalizable.  
Then demanding regularity of the solution at the horizon determines a 
discrete set of masses. The equations that describe the 
wavefunctions corresponding to motion in the $AdS$ black hole backgrounds 
have regular singular points at $0,1,{\rm horizon}$ and $\infty$ and also at
other points according to the value of $n$. In view of the discussion above, 
it might be tempting to ask for solutions which are regular at the origin 
(instead of the horizon) and which decay well enough at $\infty$, and hope 
that this would define an interesting eigenmass problem. However, a closer look
 at (\ref{new.2}) shows that the eigenvalues $m^{2}$ even if they exist are 
not guaranteed to be positive in that situation.

\section{Calculations and Results}

To leading order in $\gamma$ the equation of motion for the quadratic 
fluctuation $\eta_{0}$ of the dilaton field is \cite{EW1}
\begin{equation}
\partial_{r}(r(r^{4}-b^{4})\partial_{r}\eta_{0}) + m_{0}^{2}b^{2}r
\eta_{0} =0
\label{1.3}
\end{equation}
where one takes $b<r<\infty$ and where $m_{0}$ is the leading 
contribution to the mass in the strong coupling expansion. The eigenvalues
$m_{0}^{2}$ will provide the masses of the scalar glueball $O^{++}$ states. 
Considering first the behaviour of the solution
at infinity it is useful 
to rewrite the equation in the variable $z=b/r$ with $0<z<1$,
\begin{equation}
\frac{d}{dz}(z(\frac{1}{z^{4}}-1)\frac{d\eta_{0}}{dz})+\frac{1}{z^{3}}
\frac{m_{0}^{2}}{b^{2}}\eta_{0} =0.
\label{1.4}
\end{equation}
One wants to find normalizable wave function solutions of (\ref{1.3}) and this
fixes the behaviour at $\infty$ to be like $\eta_{0}\sim 1/r^{4}$. 
This $1/r^{4}$ behaviour at $\infty$
provides us with a Taylor expansion for $\eta_{0}$ around $z=0$ of the form 
$\eta_{0}=\sum_{n=2}^{\infty} c_{n} z^{2n}$ where to fix the overall 
normalization of $\eta_{0}$ we take $c_{1}\equiv 0$, $c_{2}=1$ and then 
obtain the recursion relation for $n\geq 2$
\begin{equation}
c_{n+1}=-\frac{c_{n}(m_{0}^{2}/b^{2})-c_{n-1}(2(n-1)(2n-3)+2(n-1))}
{(2n+2)(2n+1)-6(n+1)}.
\label{1.5}
\end{equation}

We next concentrate on the behaviour of the solutions of this equation 
near the singularity at the horizon. We will first find an expression for the 
analytic solution at the horizon and use it to reduce the order of the 
equation and show that the other independent solution is not smooth at the 
horizon. 
In order to better describe the vicinity of the horizon let us use the 
variable $\zeta = b^{2}/r^{2} -1=1/z^{2} -1$ such that the horizon is at 
$\zeta =0$\footnote{One could take $1/r -1$ as well but it turns out that 
$1/r^{2}-1$ provides a much better behaviour of the coefficients of the 
power series for the regular solution and this is important to ensure
a proper numerical treatment of the problem.}. The equation becomes
\begin{equation}
\frac{d^{2}\eta_{0}}{d\zeta^{2}}+(-\frac{1}{\zeta +1}+\frac{1}{\zeta +2}+
\frac{1}{\zeta})\frac{d\eta_{0}}{d\zeta}-\frac{m_{0}^{2}}{8b^{2}}
(-\frac{2}{\zeta +1}+\frac{1}{\zeta +2}+\frac{1}{\zeta})\eta_{0}=0
\label{1.6}
\end{equation}
where we can expand the fractional coefficients in powers of $\zeta$
and where we take a power series ansatz $\eta_{0}(\zeta)=
\sum_{n=0}^{\infty}b_n \zeta^{n}$. We obtain that the first coefficient 
$b_{0}$ is free, $b_{1}=(m_{0}^{2}/8b^{2})b_{0}$ and that the other 
coefficients 
can be determined in terms of $b_{0}$ from the recursion relation 
\begin{equation}
(n+2)^{2}b_{n+2}=\frac{m_{0}^{2}}{8b^{2}}b_{n+1}-\sum_{k=0}^{n}(k+1)(-1)^{n-k}
(\frac{1}{2^{n-k+1}}-1)b_{k+1}+
\frac{m_{0}^{2}}{8}\sum_{k=0}^{n}(-1)^{n-k}b_{k}(\frac{1}{2^{n-k+1}}-2).
\label{1.7}
\end{equation}
At the horizon this solution goes to a constant $b_{0}$ and the first
derivative $d\eta_{0}/dr=(-2/b) d\eta_{0}/d\zeta 
= (-2/b)b_{1}=(-m_{0}^{2}/4b^{3}) b_{0} \ne 0$. We will now use this 
solution to 
reduce the order of the equation and find a second linearly independent 
solution. Let $\psi$ be the analytic solution defined by the recursion 
relation (\ref{1.7}) and set $\eta_{0}=\psi\cdot g$. Inserting in 
(\ref{1.6}) we obtain a first order 
equation for the $\zeta$-derivative of $g$, 
\begin{equation}
\frac{d^2 g}{d\zeta^{2}}(\frac{dg}{d\zeta})^{-1}+
2\frac{d\psi}{d\zeta}\psi^{-1}+(-\frac{1}{\zeta +1}+\frac{1}{\zeta +2}+
\frac{1}{\zeta})=0
\label{1.8}
\end{equation}
implying that $dg/d\zeta = {\rm const.} (\zeta +1)/(\zeta (\zeta +2)\psi^{2})$.
Therefore the second solution to (\ref{1.6}) has a first derivative which 
blows up at the horizon and is not smooth there. Consequently, there are no 
smooth solutions with vanishing first derivative at the horizon. This behaviour
in the region close to the horizon is consistent with the results of 
section 2 where we have seen that the problem reduces to a Schrodinger 
problem for the harmonic oscillator perturbed by a potential of the form 
$V(x)=(-1/8)1/x^{2}$. 

We would now like to use the exact form for the series solution at the 
horizon and try to fix the overall normalization by choosing the coefficient 
$b_{0}$ in such a way that this solution matches with the Taylor expansion 
(\ref{1.5}) obtained by expanding at spatial infinity $z=0$. Of course, 
such a matching of the two
Taylor expansions over an interval will be possible only for certain values of
$m_{0}^{2}$ and these are the values for the masses. We have used these exact
expressions for the analytic expansions of the wave functions and have 
evaluated them numerically. We found that in practice this method yields strong
conditions on the allowed values of $m_{0}^{2}$ which can be found with 
arbitrarily high numerical precision. Indeed we found that for the 
allowed values of $m_{0}^{2}$ and
once we compare the values of the two Taylor expansions at one point to 
fix the coefficient $b_{0}$, the two Taylor expansions actually agree to a very
high accuracy over an entire interval thus providing an impressive test of 
the method. Small changes in the values of $m_{0}^{2}$ away from the correct 
value 
are easily detected by the mismatch they produce between the Taylor 
expansions at the 
horizon and at $\infty$. As an example we show the wave function for the 12-th
eigenvalue $m_{0}^{2}=895.8$ in Fig.1 below. We plotted the solution in 
$x=b/r$. The curve 
starting at the origin is determined from the Taylor expansion at infinity 
which we take up to $x=0.9$. We find that the power series converges extremely
well in this region. From $x=1$ we use the Taylor series from the horizon down
to $x=0.8$ where again we find a rapid convergence. As is clear from the figure
the two expansions match perfectly as expected since we are describing the
exact analytic form of the solution. We note that the radial derivative at 
the horizon is not zero (it is a factor of order one times the $x$-derivative)
and is in fact not small.
We find similar results for the other mass eigenvalues.

\vspace{0.5in}
\newpage

\begin{figure} [h]
\centerline{
\psfig{figure=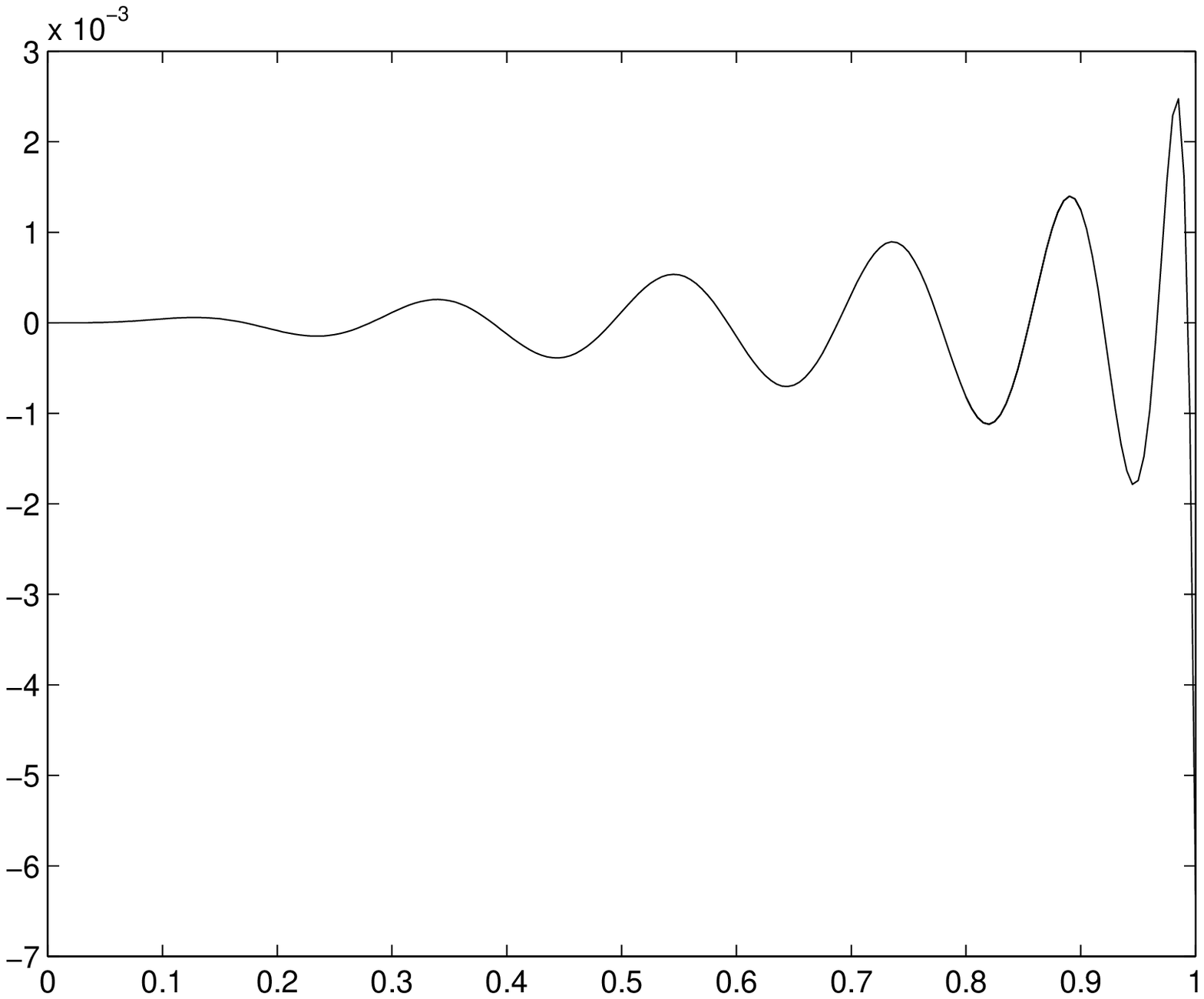,height=10cm,width=10cm}}
\end{figure}

\renewcommand{\baselinestretch}{1}

\begin{quotation}
{\small Fig1: The exact wave function for $m_{0}^{2}=895.5$.
The horizontal axis is $x=b/r$. From $x=0$ up to $x=0.9$ we use the Taylor
expansion at $\infty$. From $x=1$ down to $x=0.8$ we use the Taylor expansion 
about the horizon located at $x=1$.} 
\end{quotation}

\renewcommand{\baselinestretch}{1.5}


In Table I below we reproduce the first twelve values of $m_{0}^{2}$.


\vspace*{.5in}

\begin{center}
\begin{tabular}{|c|}
\hline
{}~~~~~SUGRA $m_{0}^{2}$
~~~~~\\
\hline
11.5877         \\
\hline
34.5270         \\
\hline
68.9750          \\
\hline
114.9104           \\
\hline
172.3312          \\
\hline
241.2366         \\
\hline
321.6265         \\
\hline
413.5009           \\ 
\hline
516.8597       \\
\hline
631.7028       \\
\hline
758.0302        \\
\hline
895.8410        \\
\hline
\end{tabular}
\end{center}

\renewcommand{\baselinestretch}{1}
{\small
\normalsize

\begin{center}
{\bf Table I}
\end{center}

\begin{quotation} 
The exact eigenvalue masses $m_{0}^{2}$ for the $O^{++}$ 
glueball in $QCD_{3}$ derived from 
supergravity. Note that 
these exact supergravity masses have been rounded to the accuracy shown.
\end{quotation}}

\renewcommand{\baselinestretch}{1.5}


The authors of \cite{COOT} used a ``shooting'' technique and numerically 
integrated the differential equation using the Taylor expansion at 
$\infty$ as an initial condition. 
To fix the values of $m_{0}^{2}$ one needs to fix the boundary condition 
at the horizon. If one uses the Neumann boundary condition $\eta_{0}^{'}=0$ 
as proposed in \cite{EW1}, one finds numerical values for $m_{0}^{2}$ that are 
in excellent agreement with the values in Table I above even though there are 
no smooth solutions satisfying that boundary condition. Although the 
dependence of the eigenmass values on the boundary condition at the horizon 
is relatively weak, this is of course not true for the wave functions since 
one is precisely discussing the first derivative at the horizon.
The fact that the actual
eigenvalues turn out to agree is interesting. It can
be explained by the fact that in the ``shooting'' technique 
$\eta_{0}^{'}(b)$ is a rapidly varying function 
of $m_{0}^{2}$ as can be seen from Fig.2 below.  We can expect that the 
discrepancy bewteen the exact mass and the one obtained using the Neumann 
boundary condition will increase with increasing masses since the exact 
boundary condition has $\eta^{'}_{0}$ at the horizon increasing with 
$m_{0}^{2}$. This point should be taken in consideration in future work on 
the subject and in particular in future studies of mass spectra from 
supergravity where one could conceivably demand a high accuracy. 

\vspace{0.5in}

\begin{figure} [h]
\centerline{
\psfig{figure=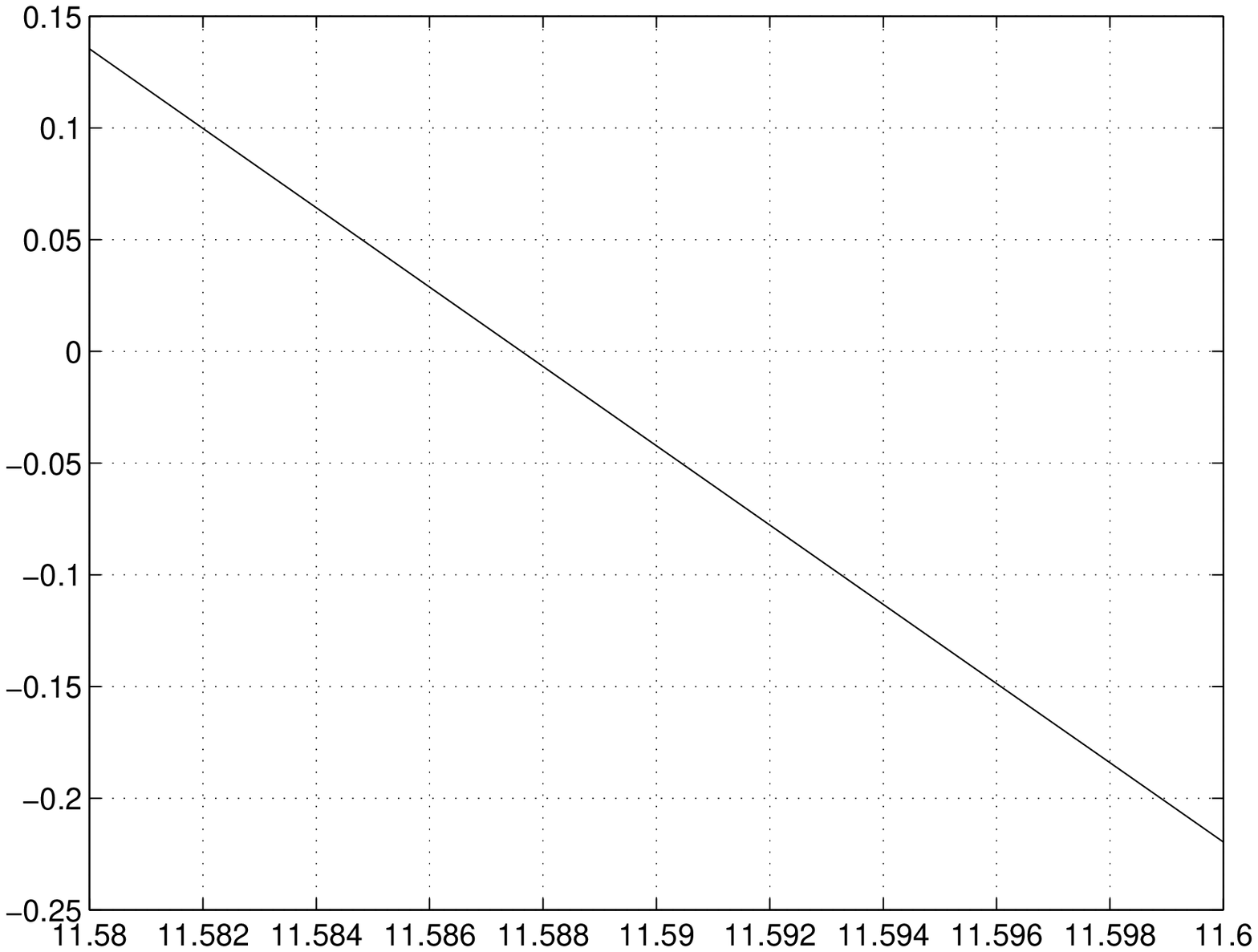,height=10cm,width=10cm}}
\end{figure}

\renewcommand{\baselinestretch}{1}

\begin{quotation}
{\small Fig2: This plot shows the dependence of the derivative
at the horizon $\frac{d\eta_{0}}{dr}(b)$ as a function of $m_{0}^{2}$ in the
``shooting'' technique. The wave function is normalized so that the first 
term in the Taylor expansion about $\infty$ is 1. In this normalization the 
exact wave function has $\frac{d\eta_{0}}{dr}=-0.03$ which would put the exact 
mass at $m_{0}^{2}$ above 11.588 whilst the Neumann boundary condition would
give $m_{0}^{2}$ below 11.588.} 
\end{quotation}

\renewcommand{\baselinestretch}{1.5}


We emphasize that our construction is based on matching the analytic forms of 
the wave functions over an extended interval. Consequently, the size of 
the error in the determination of the masses in our approach leads us to 
exclude the possibility of an exact mass formula of the form 
$m_{0}^{2}=6n(n+1)$.

To find the $O(\gamma)$ corrections to the masses of the $O^{++}$ glueballs 
one needs to study the 
equations of motion for quadratic fluctuations of the dilaton field in the 
metric background (\ref{1.1}) corrected to leading order in $\alpha^{'}$. This
correction was found in \cite{GKT} where one can also find the expression 
for $W$. One sets $m^{2}=m_{0}^{2}+\gamma m_{1}^{2}+O(\gamma^{2})$ and 
$\eta (r)=\eta_{0}(r)+\gamma \eta_{1}(r)+O(\gamma^{2})$ and perturbs 
(\ref{1.2})
about the vanishing dilaton background\footnote{Even though the dilaton is 
corrected to $O(\gamma )$ as was calculated in \cite{GKT} this does not affect 
the equation of motion for the fluctuations $\eta$ to this order.}. 
This gives the equation of motion (with $b=1$)
\begin{eqnarray}
\frac{d}{dr}(r(r^{4}-1))\frac{d\eta_{1}}{dr})+rm^{2}_{0}\eta_{1}= 
\phantom{1717171717171717171717171717171717171717171717}
\\
\nonumber
(r^{5}-r)(-\frac{300}{r^{5}}-\frac{600}{r^{9}}+\frac{1980}{r^{13}})
\frac{d\eta_{0}}{dr}+(-rm_{1}^{2}-rm_{0}^{2}(\frac{75}{r^{4}}+\frac{75}{r^{8}}
-\frac{165}{r^{12}})+\frac{405}{r^{13}}+m_{0}^{2}\frac{120}{r^{11}})\eta_{0}.
\label{1.9}
\end{eqnarray}
Normalizability and the already known behaviour of $\eta_{0}$ once 
again fix the 
behaviour of $\eta_{1}$ at $\infty$. By an analysis similar to the one we 
performed above one can show that there are no smooth solutions with 
vanishing derivative at the horizon. 
In this case the expressions for the Taylor expansions become a bit
cumbersome. 
We have calculated the first few mass
corrections $m_{1}^{2}$ by the method of matching the two Taylor expansions
for the exact solutions. Once again for the same reasons that we explained
above, we found values of $m_{1}^{2}$ which coincide with the ones obtained
by Cs\'aki et al. \cite{COOT} via the ``shooting'' technique and we will not 
repeat those values here. Therefore, we also have nothing to add to the 
physical analysis that was done in that reference and in particular on the 
comparison with the results from the lattice \cite{Lat}. 

In \cite{COOT} the authors also studied the spectrum for the $O^{--}$ 
glueball in three-dimensional QCD and the glueball mass spectra in 
four-dimensional QCD obtained from the black-hole geometry in 
$AdS_{7}\times S^{4}$ \cite{EW1}. The results of our analysis 
apply equally well to those cases. We note that in all 
these cases there is only one smooth solution at the horizon as the
indicial equation always has a double root. This is physically interesting 
since otherwise matching with the behaviour at infinity would most 
probably not put
enough restrictions on the wavefunctions and the eigenmass problem would 
likely be ill defined.

In the case of the $O^{--}$ glueball in $QCD_{3}$ one has the eigenvalue 
problem \cite{COOT}
\begin{equation}
r(r^{4}-1)\frac{d^{2}\eta_{0}}{dr^{2}}+(3+r^{4})\frac{d\eta_{0}}{dr}
+(m_{0}^{2}r-16r^{3})\eta_{0}=0.
\label{m.1}
\end{equation}
Normalizability fixes the behaviour at $\infty$ to be $\eta_{0}(r)\sim r^{-4}$ 
and the Taylor expansion at $\infty$ has the form 
$\eta_{0}(r)=\sum_{n=0}^{\infty}c_{n}x^{n+2}$ with $x=1/r^{2}$ (we set $b=1$)
and
\begin{equation}
c_{n}=\frac{(4n(n-1)+12n)c_{n-2}-m^{2}_{0}c_{n-1}}
{4(n+2)(n+1)-16+4(n+2)},
\label{m.2}
\end{equation}
with $c_{0}=1$ and  $c_{1}=-m_{0}^{2}/20$. At the horizon we use the 
variable $y=x-1$ and the ansatz $\eta_{0}(y)=\sum_{n=0}^{\infty}b_{n}y^{n}$ 
 where we obtain from (\ref{m.1})
\begin{eqnarray}
\nonumber
b_{n}= \phantom{171717171717171717171717171717171717171717171717
171717171717171717} \\
\frac{(-4(n-1)(5n-2)+m^{2}_{0}-16)b_{n-1}+(m^{2}_{0}-4(n-2)(4n-3))
b_{n-2}-4(n-3)(n-7)b_{n-3}}{8n^{2}}
\label{m.3}
\end{eqnarray}
with $b_{0}$ fixed by the Taylor expansion at $\infty$ and 
$b_{1}=(m_{0}^{2}-16)b_{0}/8$ and $b_{2}=(m^{2}_{0}-48)b_{1}/32+
m^{2}_{0}b_{0}/32$. We now find the mass eigenvalues by matching these two 
expansions. The results for $m_{0}^{2}<1000$ are shown in Table II below.


\vspace*{.5in}

\begin{center}
\begin{tabular}{|c|c|}
\hline
{}~~~~~$O^{--}$ $m_{0}$~~~~~ & $O^{--}$ ${\tilde m_{0}}$
~~~~~ \\
\hline
5.1102          &   6.11      \\
\hline
7.8234          &   9.35       \\
\hline
10.3591         &   12.39  \\
\hline
12.8375         &   15.35    \\
\hline
15.2909         &   18.28   \\
\hline
17.7280         &   21.20   \\
\hline
20.1528         &   24.09  \\
\hline
22.5718         &   26.98    \\
\hline
24.9868         &   29.88   \\
\hline
27.3998         &   32.76  \\
\hline
29.8088         &   35.64 \\
\hline
\end{tabular}
\end{center}

\renewcommand{\baselinestretch}{1}
{\small
\normalsize

\begin{center}
{\bf Table II}
\end{center}

\begin{quotation} 
Values of $m_{0}$ for the $O^{--}$ glueball in $QCD_{3}$ 
obtained from matching the exact Taylor expansions at the horizon and $\infty$
in supergravity. ${\tilde m_{0}}$ is the same mass normalized such that the
lowest $O^{++}$ mass is 4.07.
\end{quotation}}

\renewcommand{\baselinestretch}{1.5}


We observe that for the $O^{--}$ three-dimensional glueball our exact values 
for the masses differ slightly from the ones obtained in the shooting 
technique in \cite{COOT}.\footnote{However, the mass ratios are still in 
excelent agreement.}
As our first eigenvalue essentially agrees with the one in that reference we
confirm the agreement between the ratio of the lowest mass $O^{--}$ and
$O^{++}$ glueballs in $QCD_{3}$ in supergravity and on the lattice reported
in \cite{COOT}. 
For
completeness, in Fig.3 below we plot the wave function $\eta_{0}$ 
corresponding to the mass eigenvalue $m_{0}=27.3998$.

\newpage

\begin{figure} [h]
\centerline{
\psfig{figure=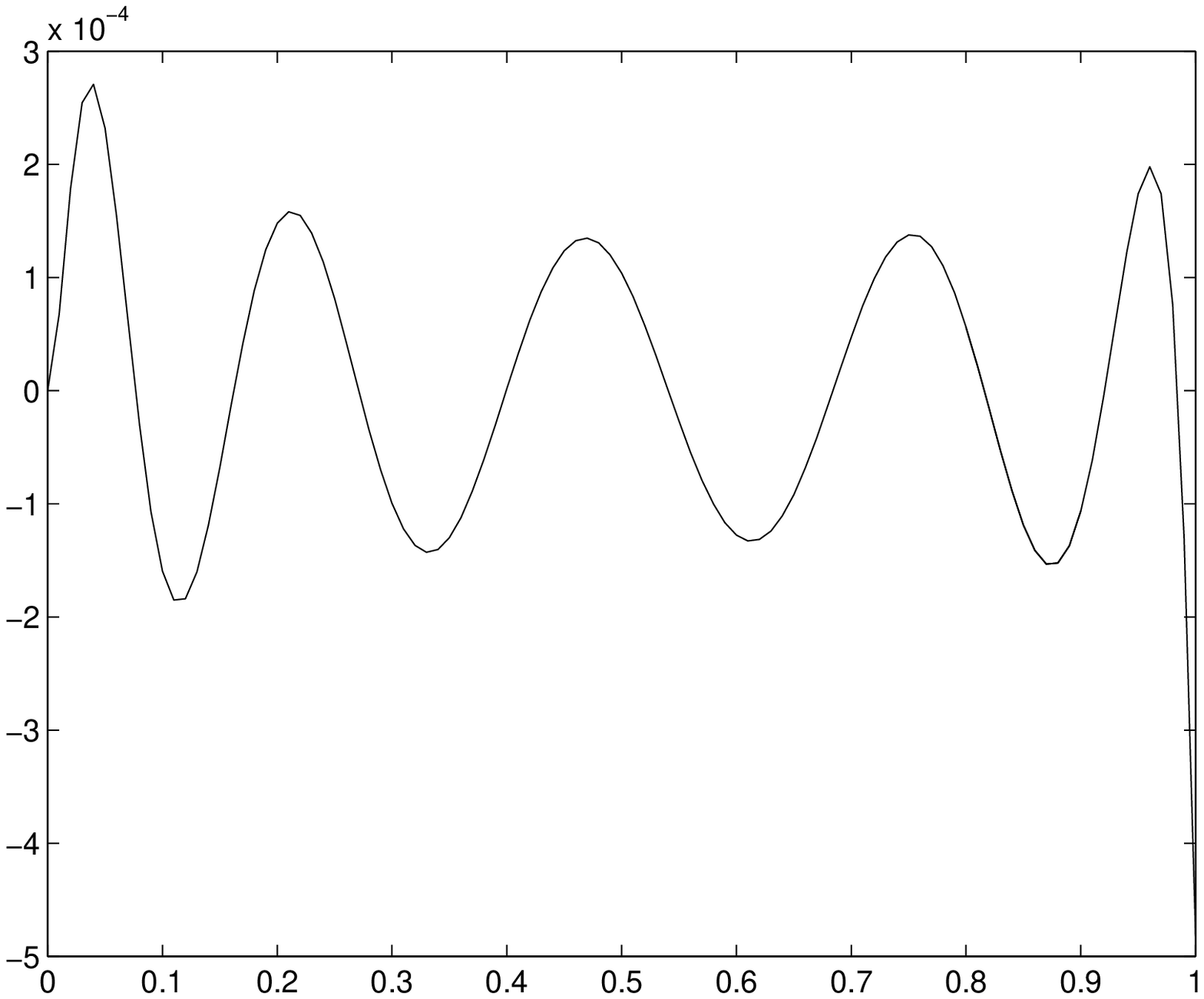,height=10cm,width=10cm}}
\end{figure}

\renewcommand{\baselinestretch}{1}

\begin{quotation}
{\small Fig3: This plot shows the wave function $\eta_{0}$ for the $O^{--}$
glueball in $QCD_{3}$ with $m_{0}=27.3998$. The plot is obtained from the 
Taylor expansions at the horizon and $\infty$ which agree perfectly.}
\end{quotation}

\renewcommand{\baselinestretch}{1.5}


Finally, we will examine the $O^{++}$ glueball in four-dimensional $QCD$.
The appropriate wave equation is in this case \cite{COOT}
\begin{equation}
(s^{7}-s)\frac{d^{2}\eta_{0}}{ds^{2}}+(8s^{6}-2)\frac{d\eta_{0}}{ds}+
s^{3}m_{0}^{2}\eta_{0} =0
\label{m.5}
\end{equation}
with $r=s^{2}$.
At $\infty$ the Taylor expansion is of the form, with $x=1/r$, 
$\eta_{0}(x)=\sum_{n=0}^{\infty}c_{n}x^{n+\frac{7}{2}}$ where $c_{0}=1$,
$c_{1}=-m_{0}^{2}/18$ and $c_{2}=m^{4}/792$. The recursion relation is
\begin{equation}
c_{n-1}=\frac{(4(n-\frac{1}{2})(n-\frac{3}{2})+2(n-\frac{1}{2}))c_{n-4}
-m_{0}^{2}c_{n-2}}{4(n+\frac{5}{2})(n+\frac{3}{2})-10(n+\frac{5}{2})}.
\label{m.6}
\end{equation}
At the horizon we use the variable $y=x-1$ and the ansatz 
$\eta_{0}(y)=\sum_{n=0}^{\infty}b_{n}y^{n}$ which gives 
\begin{eqnarray}
\nonumber
b_{n+1}= \phantom{171717171717171717171717171717171717171717171717171717
1717171717171717}\\
\frac{-2(n-2)(2n-5)b_{n-2}-2(n-1)(3+8(n-2))b_{n-1}-
(24n(n-1)-m^{2}+6n)b_{n}}{12n(n+1)+12(n+1)} \phantom{17}
\label{m.10}
\end{eqnarray}
where $b_{0}$ is fixed by matching with the Taylor expansion from $\infty$
and $b_{1}=-m_{0}^{2}b_{0}/12$ and $b_{2}=m_{0}^{2}(m_{0}^{2}-6)b_{0}/576$.
The masses we obtained are exhibited in Table III below.


\vspace*{.5in}

\begin{center}
\begin{tabular}{|c|c|}
\hline
{}~~~~~$O^{++}$ $m_{0}^{2}$~~~~~ & $O^{++}$ $m_{0}$ 
~~~~~ \\
\hline
26.9498       &   5.1913        \\
\hline
63.8820       &   7.9926        \\
\hline
114.1326      &   10.6833          \\
\hline
177.7429      &   13.3320       \\
\hline
254.7283      &   15.9602       \\
\hline
345.0944      &   18.5767   \\
\hline
448.8437      &   21.1859        \\
\hline
565.9776      &   23.7903        \\
\hline
696.4967      &   26.3912        \\
\hline
840.4013      &   28.9897      \\
\hline
997.6925      &   31.5863       \\
\hline
\end{tabular}
\end{center}

\renewcommand{\baselinestretch}{1}
{\small
\normalsize

\begin{center}
{\bf Table III}
\end{center}

\begin{quotation} 
Values of $m_{0}^{2}$ and $m_{0}$ for the $O^{++}$ glueball in $QCD_{4}$ 
obtained from matching the exact Taylor expansions at the horizon and $\infty$
in supergravity.
\end{quotation}}

\renewcommand{\baselinestretch}{1.5}


In this case our exact mass eigenvalues are also in close agreement with the
ones presented in \cite{COOT}.


\section{Conclusions}

We have examined the eigenvalue problems which feature in Witten's
generalization of the conjecture by Maldacena regarding large $N$
supersymmetric gauge theory at high temperature and strong coupling. We have
studied the eigenvalue problem through exact analytical
expansions (at both the horizon and infinity) and evaluated these 
exact expressions numerically.
We have analyzed carefully the behaviour of eigenfunctions at
the horizon and discussed the boundary conditions. We have emphasized 
the fact  
that the correct criteria for selecting the wave eigenfunctions are 
normalizability and smoothness at the horizon, have shown that no smooth 
solutions exist with vanishing derivative (Neumann boundary condition) at 
the horizon and have given a construction of such smooth solution.
Given that we are using exact analytic expressions for the various 
wave functions, our mass eigenvalues can be determined to any desired precison.

Our values for the glueball masses are in agreement with the ones found in 
\cite{COOT} and we have explained why the two techniques give identical 
results to this level of accuracy.
Since we used exact formulas for the analytic expansions of the wave function 
solutions we believe that our work reinforces the good agreement 
between various glueball spectra obtained in supergravity and on the lattice 
as was already described in \cite{COOT}. We hope that the results of this 
analysis will be of use for further studies of the conjecture. We expect
that indeed they will be necessary as soon as higher precision in the 
mass values becomes important.

{\bf{\Large Acknowledgements}}

We would like to thank Richard Easther for helpful discussions. We also wish
to thank Hirosi Ooguri for comments on the manuscript.

{\Large Note:} While the present work was in preparation, we received the 
paper \cite{COOT} which examines the same problem. In the body of the text 
we have therefore described the relation between these two studies.
After completion of this work the paper hep-th/9806128 by M. Zyskin which 
has some overlap with our paper also appeared. \\
We also received comments on the correct treatment of 
the boundary condition at the horizon by E. Witten and A. Hashimoto and
I. Klebanov (private communications).


\end{document}